\documentclass[3p,times,procedia]{elsarticle}
\usepackage{nupha_ecrc}

\volume{00}

\firstpage{1}

\journalname{Nuclear Physics A}

\runauth{S. Tripathy}

\jid{nupha}

\jnltitlelogo{Nuclear Physics A}

\usepackage{amssymb}
\usepackage[figuresright]{rotating}

\begin{document}

\begin{frontmatter}

%% Title, authors and addresses

%% use the tnoteref command within \title for footnotes;
%% use the tnotetext command for the associated footnote;
%% use the fnref command within \author or \address for footnotes;
%% use the fntext command for the associated footnote;
%% use the corref command within \author for corresponding author footnotes;
%% use the cortext command for the associated footnote;
%% use the ead command for the email address,
%% and the form \ead[url] for the home page:
%%
%% \title{Title\tnoteref{label1}}
%% \tnotetext[label1]{}
%% \author{Name\corref{cor1}\fnref{label2}}
%% \ead{email address}
%% \ead[url]{home page}
%% \fntext[label2]{}
%% \cortext[cor1]{}
%% \address{Address\fnref{label3}}
%% \fntext[label3]{}

%% Instructions from Editor: Please use the following \dochead only in the preprint version (e-print arXiv etc.); 
%% use empty \dochead{} when submitting to Nuclear Physics A!
\dochead{XXVIIth International Conference on Ultrarelativistic Nucleus-Nucleus Collisions\\ (Quark Matter 2018)}
%\dochead{}
%% Use \dochead if there is an article header, e.g. \dochead{Short communication}
%% \dochead can also be used to include a conference title, if directed by the editors
%% e.g. \dochead{17th International Conference on Dynamical Processes in Excited States of Solids}

\title{Energy dependence of $\phi$(1020) production at mid-rapidity in pp collisions with ALICE at the LHC}

%% use optional labels to link authors explicitly to addresses:
%% \author[label1,label2]{<author name>}
%% \address[label1]{<address>}
%% \address[label2]{<address>}

\author{Sushanta Tripathy (for the ALICE Collaboration)}

\address{Discipline of Physics, Indian Institute of Technology Indore, Simrol, Madhya Pradesh - 453552, India}

\begin{abstract}
Hadronic resonances are unique tools to investigate the interplay of
re-scattering and regeneration effects during the hadronization phase in heavy-ion
collisions. Measurements in small collision systems provide a necessary baseline for heavy-ion data, help to tune pQCD inspired event generators and give insight into the search for the onset of collective effects. 
As the $\phi$ meson has a longer lifetime compared to other
resonances, it is expected that its production would be much less affected by
regeneration and re-scattering processes. We report on measurements of $\phi$ meson production in minimum bias pp collisions at different beam energies and as a
function of charged particle multiplicity with the ALICE detector at the LHC. The results include the transverse
momentum $(p_{\mathrm{T}})$ distributions of $\phi$ as well as the particle yield ratios. Finally, we have also studied the $\phi$ effective strangeness content by comparing our results to theoretical calculations.

\end{abstract}

\begin{keyword}
%% keywords here, in the form: keyword \sep keyword
Resonances; Strangeness Enhancement; Particle production mechanism; Hadronic phase
%% MSC codes here, in the form: \MSC code \sep code
%% or \MSC[2008] code \sep code (2000 is the default)

\end{keyword}

\end{frontmatter}

%%
%% Start line numbering here if you want
%%
%\linenumbers

%% main text
\section{Introduction}
\label{int}
Resonances are important tools to probe the hadronic phase formed in heavy-ion collisions due to their short lifetimes. $\phi$ is a hidden strange vector meson, whose mass is similar to that of proton and $\Lambda$. The lifetime of $\phi$ (46.3 fm/{\it c}) is longer compared to other hadronic resonances as well as the fireball produced in heavy-ion collisions. Thus it is expected that $\phi$ meson may not go through the re-generation and re-scattering processes~\cite{regen} occurring during the hadronic phase. Being the $\phi$ formed by an $\mathrm{s\bar{s}}$ quark pair with zero net-strangeness content, measurements of $\phi$ meson production can help seeding light on strangeness production mechanisms. Furthermore, the study of $\phi$ in pp collisions helps to search for the onset of collectivity in small systems and provide a necessary baseline for heavy-ion collisions. 

This article focuses on the study of the energy and charged particle multiplicity dependence across different colliding systems of $\phi$ production, by comparing results obtained in pp collisions at $\sqrt{s}=2.76$~\cite{2017}, $5.02$, $7$~\cite{2012}, $8$ and $13$~TeV at the LHC. In particular, $p_\mathrm{T}$ spectra at different energies as well as $p_\mathrm{T}$-integrated particle ratios to long-lived hadrons are compared. In this paper, we aim to address the following three questions: is there any dependence of $\phi$ production in pp collision on collisions energy or on charged particle multiplicity, do re-scattering and regeneration processes affect the production of the long-lived $\phi$ in high multiplicity pp collisions and, finally, how does the effective strangeness content play a role in strangeness production mechanism?

%% The Appendices part is started with the command \appendix;
%% appendix sections are then done as normal sections
%% \appendix

\section{$\phi$ meson reconstruction and $p_\mathrm{T}$ spectra}
\label{rec}

The $\phi(1020)$ is reconstructed through an invariant mass analysis at mid-rapidity ($|y|<$ 0.5) of its hadronic decay channel~\cite{2017,2012} i.e. K$^{+}$K$^{-}$ (Branching ratio: 48.9\%)~\cite{pdg} using Inner Tracking System (ITS), Time Projection Chamber (TPC) and Time of Flight (TOF) detector of ALICE. The yields of $\phi(1020)$ in different $p_{\mathrm{T}}$ intervals are obtained by subtracting the combinatorial background from the unlike-sign charged particle invariant mass distribution. The event mixing and like-sign techniques are used to estimate the combinatorial background. After combinatorial background subtraction a residual background remains which arises mainly due to mis-identified particle decay products or from other sources of correlated pairs (e.g. mini-jets). The $\phi(1020)$ peak is fitted with a Voigtian function, which is a convolution of Breit-Wigner and Gaussian functions~\cite{2017,2012}. For the cases when the combinatorial background shows large statistical fluctuation, the $\phi$ peak is fitted without any combinatorial background subtraction. In different $p_{\mathrm{T}}$ intervals, raw yields are obtained from the residual background subtracted signal distributions. The raw yields are then corrected for the detector efficiency $\times$ acceptance and the branching ratios to determine the final $p_{\mathrm{T}}$ spectrum. The estimation of charged particle multiplicity classes is performed using V0 detector.

%Figure~\ref{fig2} shows the $p_{\mathrm{T}}$  spectra of $\phi$ mesons in pp collisions at $\sqrt{s}$ = 5.02 and 13 TeV in different multiplicity classes. The lower panels show the ratio of the $p_{\mathrm{T}}$ spectra to the full 0-100\% (INEL$>$0) $p_{\mathrm{T}}$ spectrum. The ratios increase at low-$p_{\mathrm{T}}$ and remain flat at high-$p_{\mathrm{T}}$, indicating that the bulk production increases with charged particle multiplicity.

%\begin{SCfigure}
%\caption[]{Invariant-mass distribution for the $\phi$ in pp collisions at $\sqrt{s}$ = 5.02 TeV in the $p_{\mathrm{T}}$ range 0.5 $<$ $p_{\mathrm{T}}$ $<$ 0.7 GeV/c in V0M Multiplicity class VII. Left: the unlike-charge invariant-mass distribution with combinatorial backgrounds. Right: Invariant-mass distribution after subtraction of the mixed-event background with a fit to describe the peak of the $\phi$ and the residual background.}
%\centering
%\includegraphics[height=10em]{2018-May-09-Inv_mass_30_40.pdf}
%\includegraphics[height=10em]{2018-May-09-Inv_mass_fit_30_40.pdf}
%\label{fig1}
%\end{SCfigure}

%\begin{figure}[ht!]
%\centering
%\includegraphics[height=15.3em]{2018-May-09-PhipTspectra.pdf}
%\includegraphics[height=15.1em]{2017-Oct-01-phi_spectra_rmb_linx.pdf}
%\caption[]{$p_{\mathrm{T}}$  spectra of $\phi$ mesons in pp collisions at $\sqrt{s}$ = 5.02 TeV (left) and 13 TeV (right) in different multiplicity classes. In the bottom panels of the figure the ratios of the $p_{\mathrm{T}}$ spectra to the full 0-100\% (INEL$>$0) $p_{\mathrm{T}}$ spectrum is reported.}
%\label{fig2}
%\end{figure}

\section{Results and Discussion}
\label{rec}

\begin{figure}[ht!]
\centering
\includegraphics[height=11.9em]{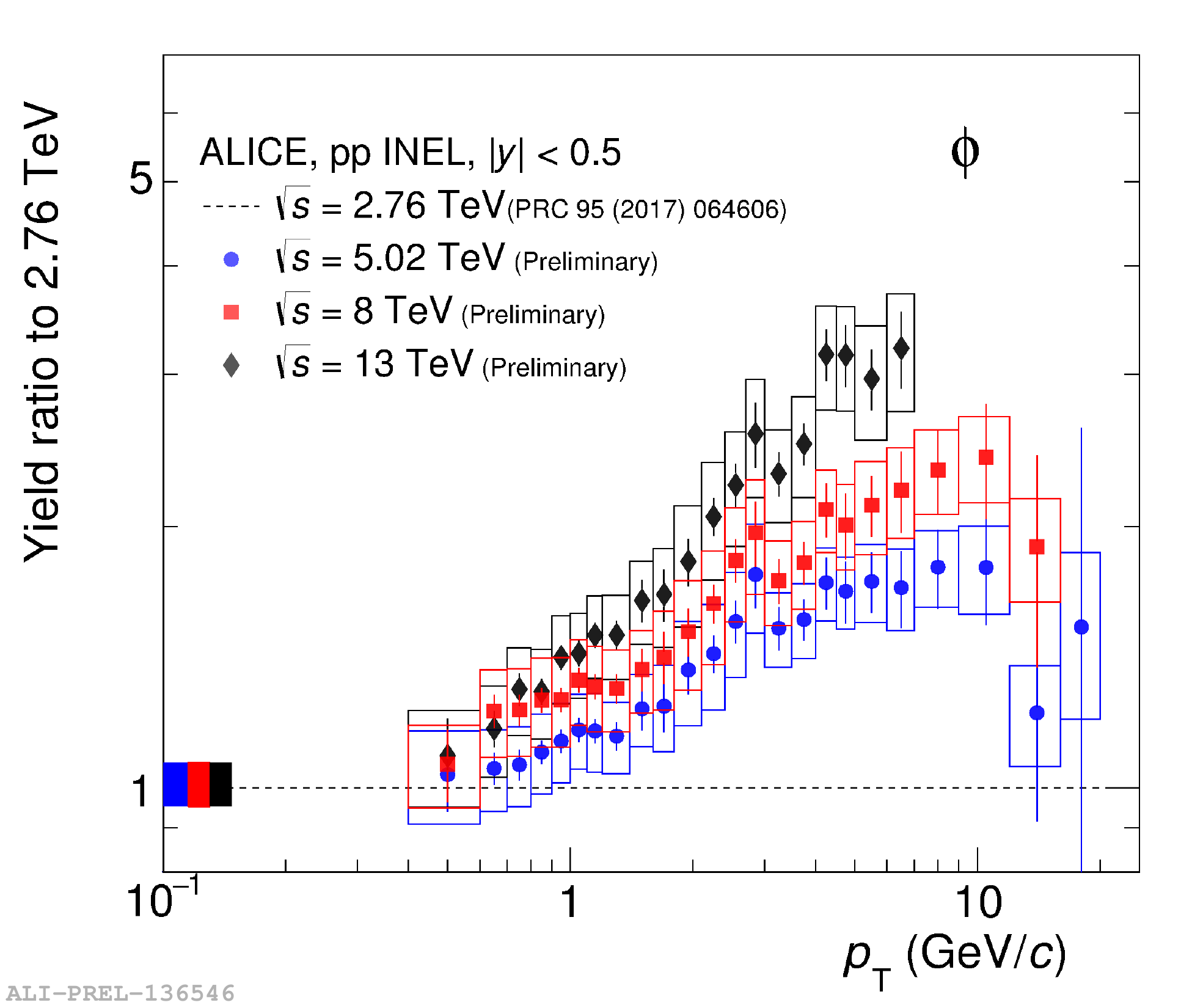}
\includegraphics[height=11.9em]{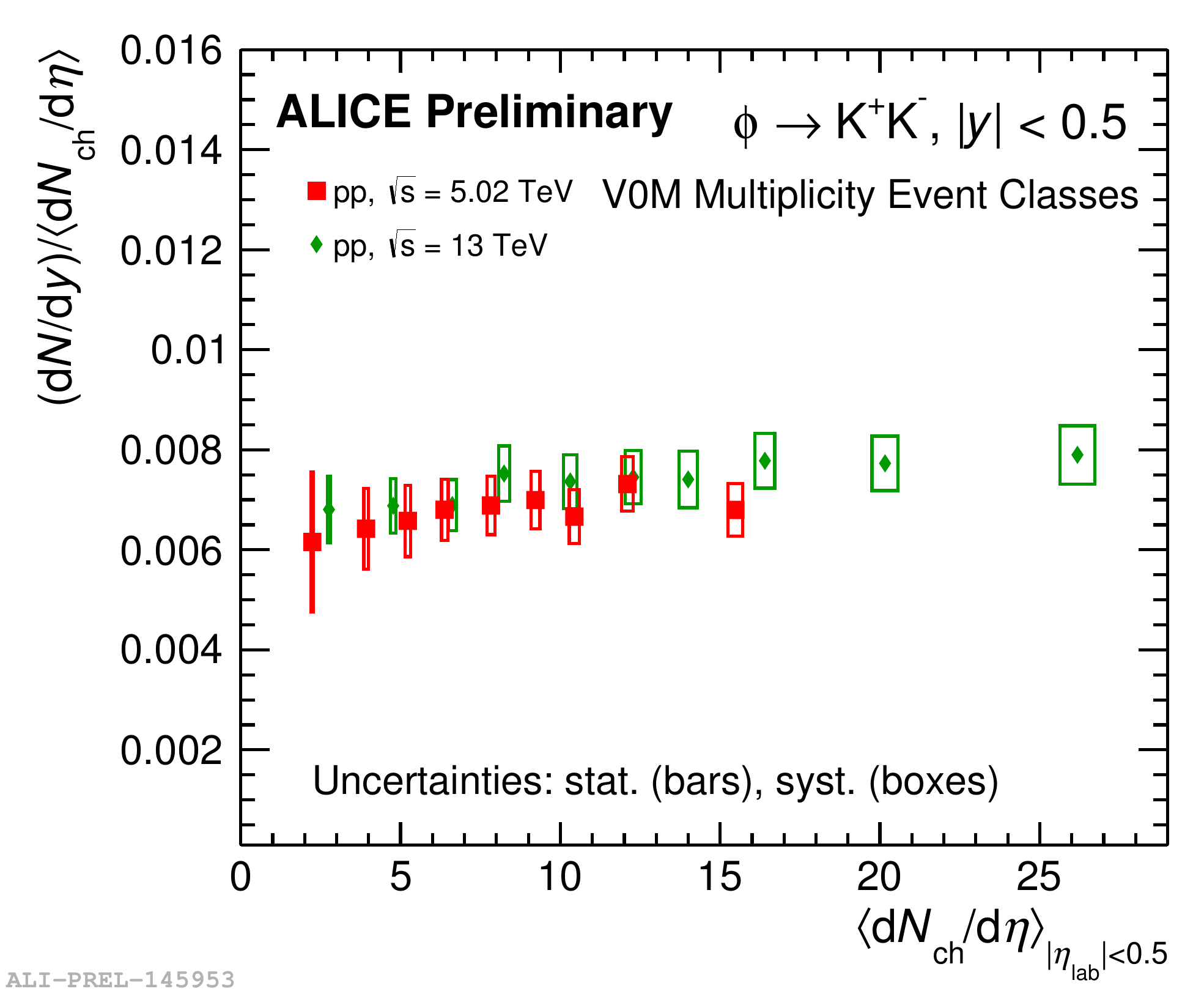}
\caption[]{Left: Ratios of $p_{\mathrm{T}}$ spectra of $\phi$ in inelastic pp collisions at various center of mass energies to the spectrum obtained in pp collisions at $\sqrt{s}$ = 2.76 TeV. Right: ($\mathrm{d}N/\mathrm{d}y$)/$\langle \mathrm{d}N_{\mathrm{ch}}/\mathrm{d}\eta\rangle$ for $\phi$ as a function of average charged particle multiplicity in pp collisions at $\sqrt{s}$ = 5.02 and 13 TeV. Statistical uncertainties are represented by bars and systematic uncertainties are represented by boxes.}
\label{fig3}
\end{figure}

\begin{figure}[ht!]
\centering
\includegraphics[height=12.2em]{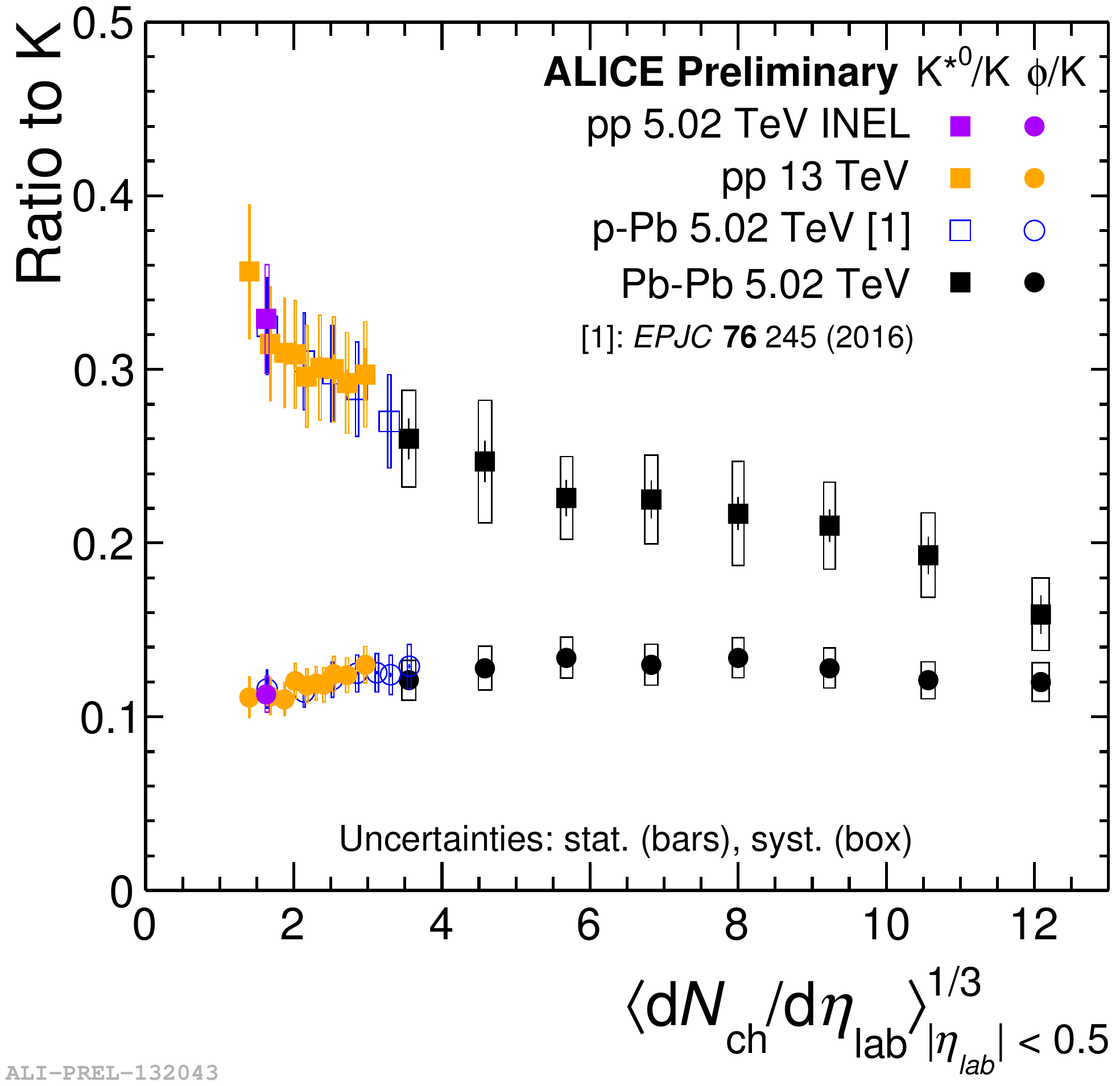}
\includegraphics[height=12.6em]{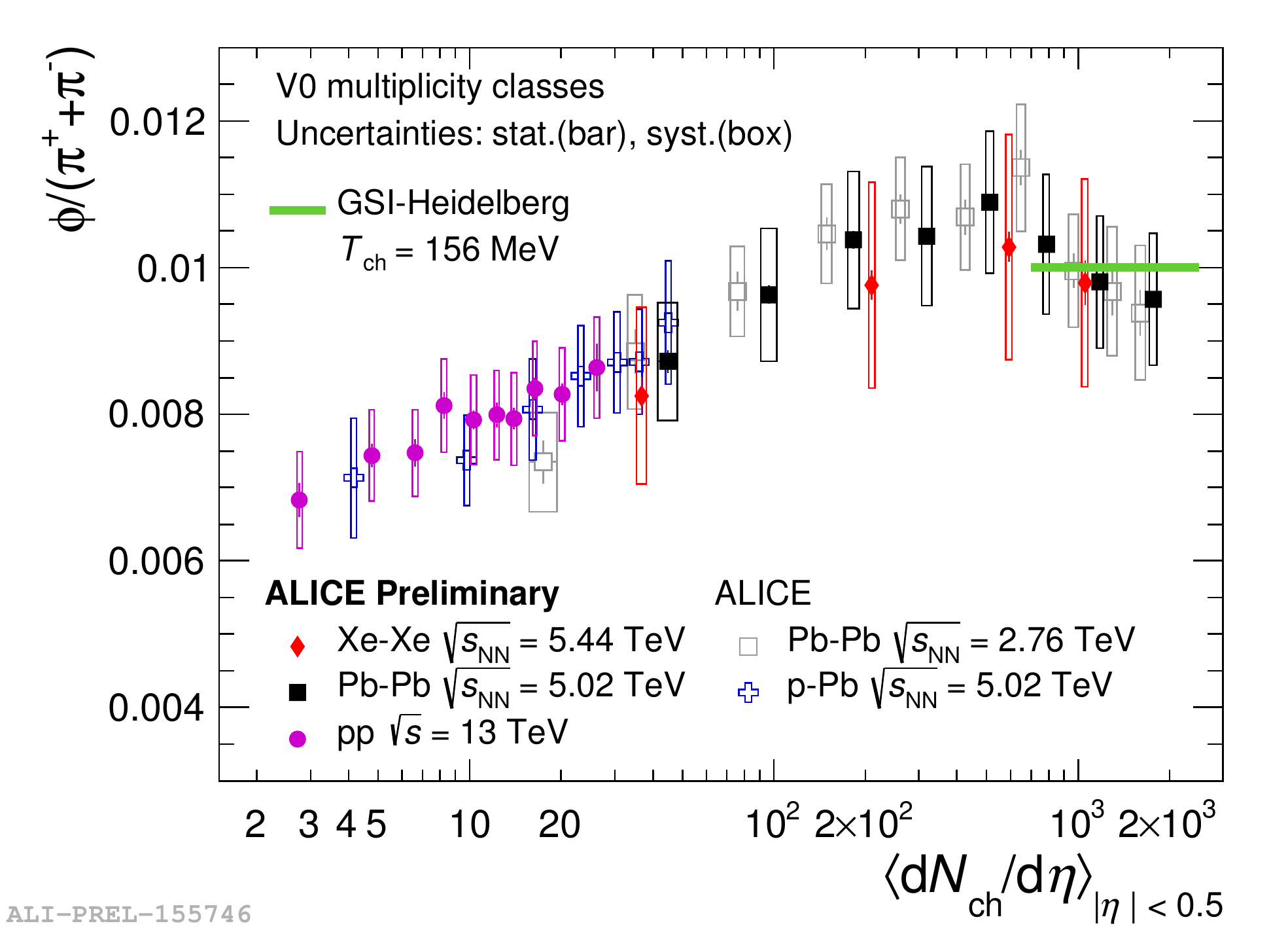}
\includegraphics[height=12.2em]{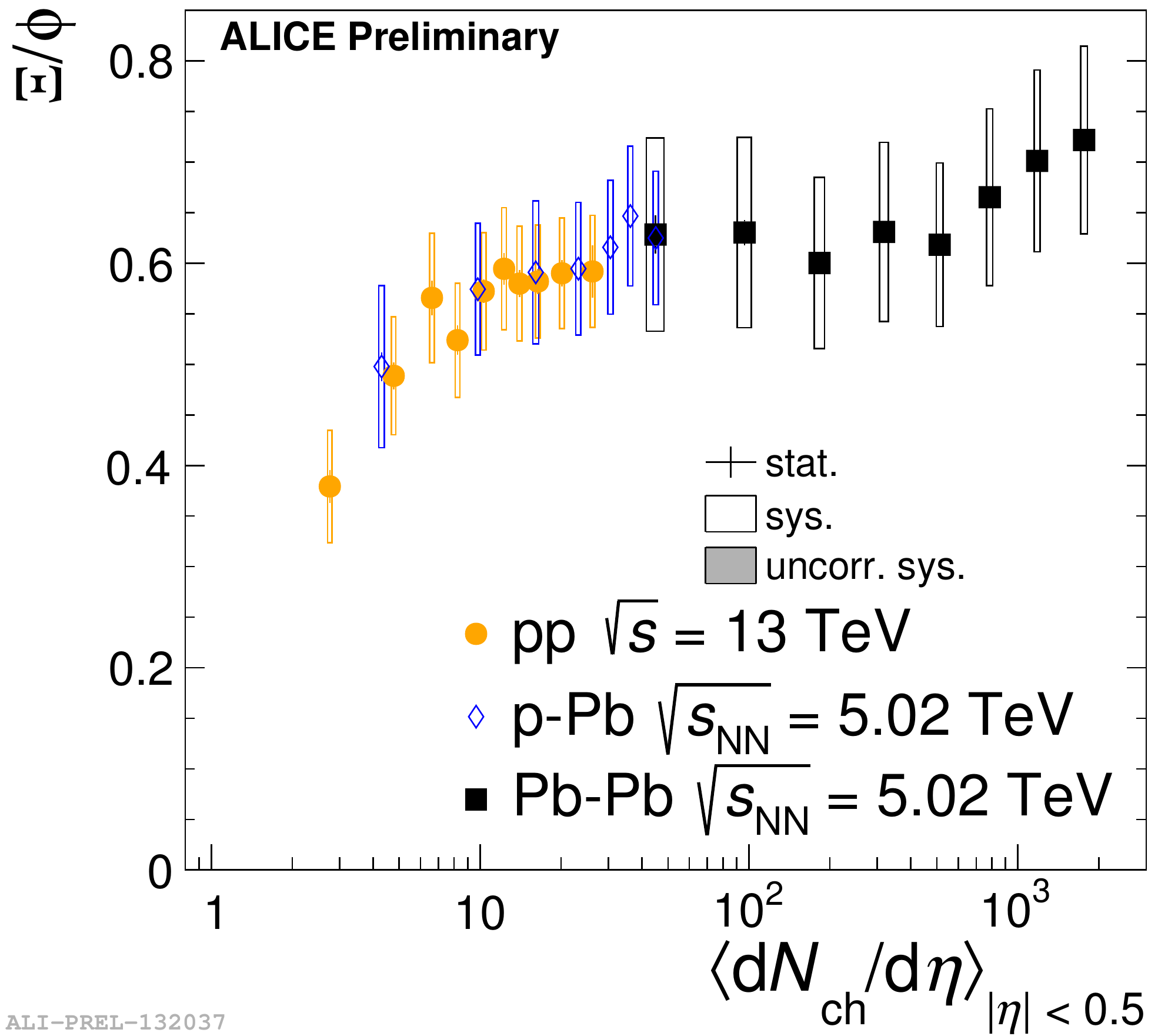}
\caption[]{Ratios of $p_{\mathrm{T}}$-integrated yield of $\phi$ and K$^{*0}$ to K (top left), $\phi$ to $\pi$ (top right) and $\Xi$ to $\phi$ (bottom) as a function of charged particle multiplicity in different collision systems for different center-of-mass energies. Here, K$^{*0}$ and $\Xi$ are reconstructed at mid-rapidity ($|y|< \rm{0.5}$) and the reconstruction of K$^{*0}$ and $\Xi$ are described in detail in~\cite{2017,Adam:2015vsf}. The charged particle multiplicity is estimated in V0 multiplicity classes.}
\label{fig4}
\end{figure}

Figure~\ref{fig3} (left) shows ratios of $\phi$ $p_{\mathrm{T}}$ spectra in inelastic pp collisions at various center of mass energies to the spectrum obtained in pp collisions at $\sqrt{s}$ = 2.76 TeV. These ratios indicate that from 1-2 GeV/$\it c$, the spectra increases as a function of collision energy while the bulk production at low $p_{\mathrm{T}}$ does not strongly depend of collision energy. The right panel of Figure~\ref{fig3} shows the $\phi$ normalised ($\mathrm{d}N/\mathrm{d}y$)/$\langle \mathrm{d}N_{\mathrm{ch}}/\mathrm{d}\eta\rangle$ value as a function of average charged particle multiplicity in pp collisions at $\sqrt{s}$ = 5.02 and 13 TeV. It remains flat and independent of collision energy, which suggests that the event multiplicity drives the particle production, irrespective of collision energy. The top left panel of Figure~\ref{fig4} shows the ratio of $\phi$ and K${^{*0}}$ to kaons as a function of the charged particle multiplicity. As the lifetime of K$^{*0}$ is $\simeq$10 times shorter compared to $\phi$, it is expected that K${^{*0}}$ would be affected by the re-generation and re-scattering processes. A decreasing trend of K$^{*0}$/K ratio is observed, which suggests the re-scattering dominates over regeneration process. As expected, the $\phi$/K ratio remains fairly flat indicating thus either the regeneration and re-scattering are balanced or, given its longer lifetime, $\phi$ decays after the hadronic phase and is not affected by re-scattering and regeneration. The top right and bottom panels of Figure~\ref{fig4} show $\phi$/$\pi$ and $\Xi$/$\phi$ ratios as a function of $\langle \mathrm{d}N_{\mathrm{ch}}/\mathrm{d}\eta\rangle$ respectively. In the $\phi/\pi$ ratio, the production of $\phi$ in large systems (Pb--Pb and Xe--Xe collisions) is well described by a grand-canonical thermal model (GSI-Heidelberg)~\cite{Stachel:2013zma}, while for small systems (pp and p--Pb collisions) the increase of the $\phi/\pi$ ratio with multiplicity is in contrast to the expectation from strangeness canonical suppression (not shown)~\cite{Vislavicius:2016rwi}. This behavior favors the non-equilibrium production of $\phi$ and/or strange particles. The $\phi/\mathrm{K}$ and $\Xi/\phi$ ratios remain fairly flat across a wide multiplicity range. Comparing the $\phi$ with particles with strange quark content =1 (K) or =2 ($\Xi$) we observe that the $\phi$ behaves similarly to particles with open strangeness. In addition, there is a multiplicity dependence for the $\Xi/\phi$ ratio observed, particularly at low multiplicities.

\section{Summary}
\label{sum}
The ALICE collaboration has studied $\phi$ production as a function of collision energy and charged particle multiplicity. The event multiplicity seems to be driving the $\phi$ production, irrespective of collision energy. The $\phi$/K ratio remains rather flat, which indicates that either regeneration and re-scattering are balanced or that the $\phi$ decays after the hadronic phase and is not affected by re-scattering and regeneration for Pb--Pb collisions. The trend of $\phi$/$\pi$ in small collision systems is inconsistent with simple canonical suppression. We have observed that the $\phi$ meson shows a similar behavior to that of particles with open strangeness.
\\
\\
%\section{Acknowledgements}
ST acknowledges the financial support by DST-INSPIRE program of Government of India.

%% References
%%
%% Following citation commands can be used in the body text:
%% Usage of \cite is as follows:
%%   \cite{key}         ==>>  [#]
%%   \cite[chap. 2]{key} ==>> [#, chap. 2]
%%

%% References with BibTeX database:

%\bibliographystyle{elsarticle-num}
%\bibliography{<your-bib-database>}

%% Authors are advised to use a BibTeX database file for their reference list.
%% The provided style file elsarticle-num.bst formats references in the required Procedia style

%% For references without a BibTeX database:

\end{document}